\documentclass[aps,pra,preprintnumbers,amsmath,amssymb,floatfix,superscriptaddress,twocolumn]{revtex4}
\usepackage{txfonts}
\usepackage{graphicx} 
\usepackage{bm}       
\usepackage{color}
\usepackage{subfigure}

\newcommand{\tr}[1]{\text{Tr}}
\newcommand{\ket}[1]{|#1\rangle}

\newcommand{\sprod}[2]{\langle#1|#2\rangle}
\begin{document}
\title{The role of quantum non-Gaussian distance in entropic uncertainty relation}
\author{Wonmin Son}
\email{sonwm@physics.org}
\affiliation{Department of Physics, Sogang University, Mapo-gu, Shinsu-dong, Seoul 121-742, Korea}
\date{\today}

\begin{abstract}
Gaussian distribution of a quantum state with continuous spectrum is known to maximize the Shannon entropy at a fixed variance. Applying it to a pair of canonically conjugate quantum observables $\hat x$ and $\hat p$, quantum entropic uncertainty relation can take a suggestive form, where the standard deviations $\sigma_x$ and $\sigma_p$ are featured explicitly. From the construction, it follows in a transparent manner that: (i) the entropic uncertainty relation implies the Kennard-Robertson uncertainty relation in a modifed form, $\sigma_x\sigma_p\geq\hbar e^{\cal N}/2$; (ii) the additional factor ${\cal N}$ quantifies the quantum non-Gaussianity of the probability distributions of two observables; (iii) the lower bound of the entropic uncertainty relation for non-gaussian continuous variable (CV) mixed state becomes stronger with purity. Optimality of specific non-gaussian CV states to the refined uncertainty relation has been investigated and the existance of new class of CV quantum state is identified.
\end{abstract}

\maketitle

\section{Introduction} 
 
In quantum mechanics, the ``uncertainty principle" states that there is a fundamental limit to our ability to assign well defined values to two complementary observables, such as the position $\hat x$ and the momentum $\hat p$. The original heuristic argument that such a limit should exist was given by Werner Heisenberg in 1927 \cite{Heisenberg27}. Shortly after Heisenberg's seminal paper, Kennard and Robertson derived the well-known inequality ($\hbar=1$)
\begin{equation}
\sigma_x\sigma_p \ge 1/2,
\label{HUR}
\end{equation}
valid when $[\hat x,\hat p]=i$ \cite{Kennard27, Robertson29} and relating the standard deviations $\sigma_x$, $\sigma_p$ of the position and the momentum distribution, respectively. For simplicity, and with an abuse in terminology, we shall refer to the relation as Heisenberg uncertainty relation. Since these initial insights, uncertainty relations (URs) have become a central topic in studies of quantum mechanics. Later, the uncertainty principle has been formulated in terms of information theoretical quantities \cite{Hirschmann57,Beckner75,Biaynicki75}. 
In particular, the {\it entropic uncertainty relation} (EUR) for $\hat x$ and $\hat p$ reads
\begin{equation}
	H(X)+H(P)\geq\ln(\pi e),
\label{EUR}
\end{equation}
where $H$ is Shannon's differential entropy \cite{Shannon1948}, and $X$,$P$ are the classical random variables associated with the measurement of position and momentum, respectively. We recall that the explicit expression for $H(X)$ is 
\begin{equation}
\label{ENT}
H(X)=-\int dx\, {\cal P}(X=x) \ln {\cal P}(X=x),
\end{equation}
where ${\cal P}(X=x)$ is the probability density for the outcome $x$, and an analogous expression holds for $H(P)$. Heisenberg UR in Eq.\eqref{HUR} and entropic UR in Eq.\eqref{EUR} can be seen as two different formalizations of the uncertainty principle. Initially, the latter has been known to be stronger \cite{Uffink88} while it turned out that they characterize quantum states in a different manner \cite{Baek14}. However, a proper comparison of the relations for the case of CV state has not been made before and the comparison of UR on CV states is the central topic in this work. Furthermore, it is also interesting to note that a proper mathematical formalization of the case dependent URs is still under debate and their operational meanings are getting to be clearer with recent active investigations \cite{Ozawa01,Branciard13,Bush13,Buscemi14}. 

While it is known that Eq.(\ref{HUR}) can be used to derive Eq.(\ref{EUR}), we further show that a very simple relationship exists between the two in this paper. Specifically, we shall show that the maximum entropy principle allows one to rewrite the entropic UR as a refinement of Heisenberg's UR, which explicitly features the deviation of the position and the momentum probability densities from appropriate reference Gaussians. To the best of our knowledge, this simple and yet powerful observation has not been noticed anywhere before. This paper is organised as follows: In Section II we recall Shannon's application of the maximum entropy principle, which we use in Section III to establish an exact relationship between the entropic UR and Heisenberg's uncertainty. In Section IV we discuss various examples of non-Gaussian states, and in Section V we draw our conclusions.

\section{Maximum entropy principle and negentropy}
The principle of maximum entropy states that the most conservative guess for the (unknown) probability distribution of a certain random variable can be obtained by maximizing the Shannon entropy under the known constraints  \cite{Shannon1948}. Shannon applied the principle to the case of a continuous variable with fixed standard deviation $\sigma$ and showed that the differential entropy was maximized by the Gaussian distribution 
\begin{equation}
p(x)=\frac{1}{\sigma\sqrt{2\pi}}e^{-(x-\langle x\rangle)^2/2\sigma^2}.
\end{equation}
 We shall indicate the entropy of such distribution $H_G(\sigma)=\ln(\sqrt{2\pi e}\sigma)$, as it is a clear function of the variance only. Conversely, this implies that the following quantity is nonnegative:
\begin{equation}
\label{eq:NGD}
{\cal J}(X)=H_G(\sigma_x)-H(X),
\end{equation}
also known as {\it neg-entropy} \cite{Schrodinger44}. This can be used to quantify the deviation between a probability distribution and its Gaussian reference (i.e. the Gaussian distribution with the same variance), since clearly ${\cal J}(X)=0$ if and only if  $X$ is Gaussian distributed and ${\cal J}(X)> 0$ for any non-gaussian $X$. ${\cal J}(X)$ can thus be taken as a possible quantification of {\it non-Gaussianity} of the classical random variable $X$. 

\section{Application to entropic uncertainty relation}
In the previous section, a formal definition and its properties of neg-entropy have been presented. The properties natrually provide casual proof that, for any random variable $X$, the Shannon entropy of an arbitrary profiled CV state has an upper bound as $$H(X)\leq H_G(\sigma_x)=\ln(\sqrt{2\pi e}\sigma_x).$$ At the same time, for a single variable, no lower bound to $H(X)$ exists, and arbitrary negative values can be achieved; the upper bound can instead be used to define the relative entropy or Kullback-Leibler divergence \cite{Jaynes63}. An equivalent statement is that ${\cal J}(X)\geq0$, but ${\cal J}(X)$ has no upper bound \cite{Comon94}. Contrarily, due to the entropic uncertainty relation in Eq.~\eqref{EUR}, a lower bound does however exist when considering the combined entropies of two canonically conjugate quantum variables, which will be discussed shortly. 

In order to quantify the non-Gaussianity in a quantum state, one should use the combined neg-entropy. For the case of quantum state, non-Gaussian properties in more than one quadrature are needed to be tested as like entropic UR in (\ref{EUR}). Using the quantification, our main result can be obtained simply by rewriting Eq.~\eqref{EUR} in terms of the combined neg-entropies. Noting that $H(X)=H_G(\sigma_x)-{\cal J}(X)$ (and the analogue expression for $P$), we have $H_G(\sigma_x)+H_G(\sigma_p)\geq\ln(\pi e)+{\cal J}(X)+{\cal J}(P)$ from (\ref{EUR}) and (\ref{eq:NGD}). Taking exponentials on both sides and performing some simple algebra, we obtain
\begin{equation}
	\sigma_x\sigma_p\geq\frac{1}{2} e^{{\cal J}(X)+{\cal J}(P)}.
\label{main}
\end{equation} 
Eq.~\eqref{main} is equivalent to entropic UR, but it has instructive form with modification. It shows explicitly that entropic UR implies the Heisenberg uncertainty relation, and that a {\it necessary} condition for a state to saturate the latter is to have Gaussian marginals. Hence, in continuous variables there is a hierarchy between the two types of uncertainty relation, differently from the finite-dimensional case \cite{Deutsch83, Uffink88, Baek14}. More importantly, we identify that the ``gap'' between the two uncertainty relations is exactly quantified by the (exponentiated) non-Gaussianity of the two variables. In fact, Eq.~\eqref{main} can also be seen as an {\it upper bound} to the combined neg-entropies of $X$ and $P$  for given variances $\sigma_x,\sigma_p$:
\begin{equation}
\label{eq:Neg_bound}
{\cal J}(X)+{\cal J}(P) \leq \ln(2\sigma_x\sigma_p).
\end{equation}
We remark that the combination ${\cal J}(X)+{\cal J}(P)$ depends on the quantum state under consideration, as well as on the choice of the canonically conjugate observables to be measured. This is in sharp contrast to typical studies of {\it quantum non-Gaussianity} \cite{Genoni08,Mandilara12}, as they are concerned with quantities that depend solely on the quantum state $\rho$ but not measurements. For the purpose of the present study, an interesting hybrid of classical and quantum concepts appears to provide the most relevant quantification of non-Gaussianity. For simplicity, from now on, we shall refer to the quantity $${\cal N}\equiv{\cal J}(X)+{\cal J}(P)$$ simply as {\it quantum neg-entropy} or {\it gaussian distance} as in the spirit of classical non-Gaussianity. In the following section, we will investigate the quantative behavior of the property for the case of well-known non-Gaussian CV quantum states. 

\section{Specific quantum state examples}
\begin{figure}[t]
\begin{center}
\subfigure[$~$Photon number state. The difference ${\cal B}-{\cal N}$ becomes larger as the photon number increased.]{
   \includegraphics[scale =0.46] {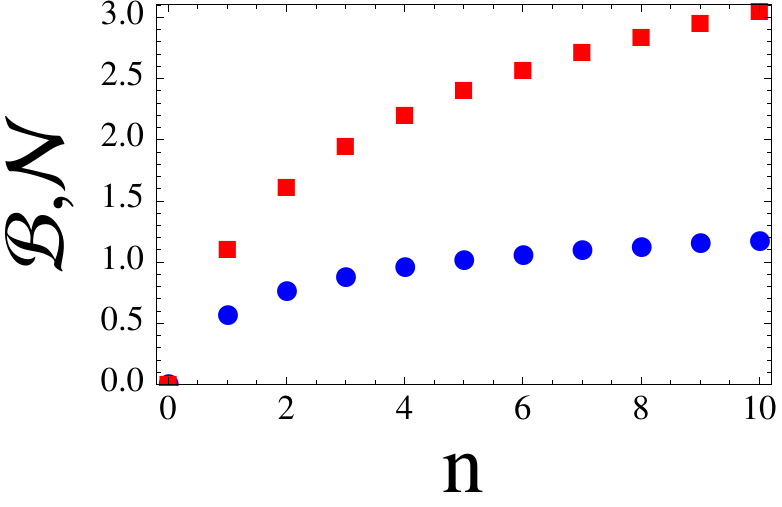}
   \label{fig:subfig1}
 }
\quad
\subfigure[$~$Possion/Laplace state. ${\cal B}$ and ${\cal N}$ remain constant for any value of standard deviation.]{
   \includegraphics[scale =0.45] {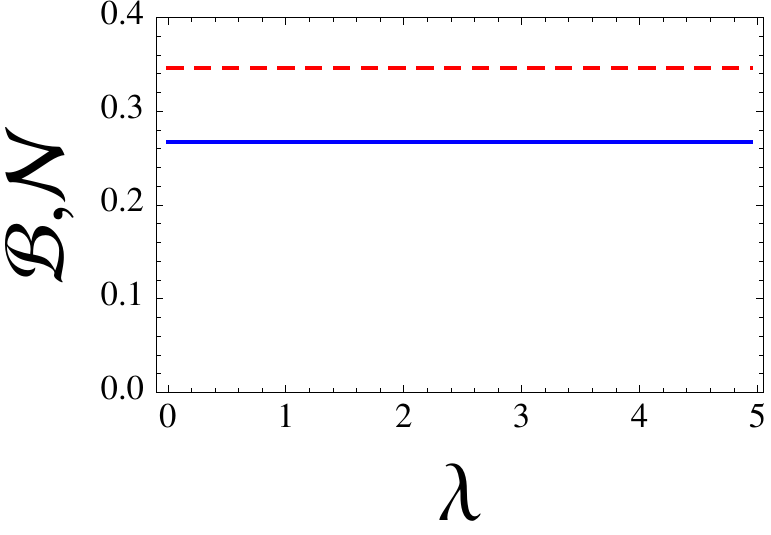}
   \label{fig:subfig2}
 }
\subfigure[$~$Photon added coherent state. The bound becomes zero as the state approaches to gaussian state.]{
   \includegraphics[scale =0.45] {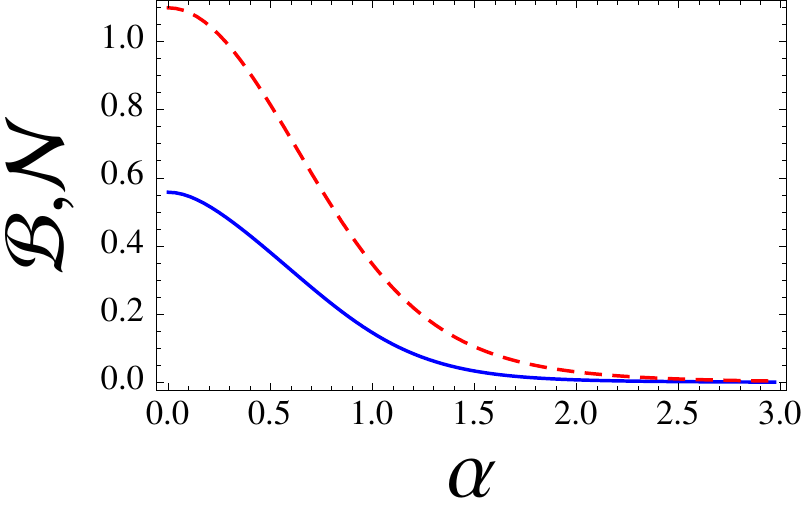}
   \label{fig:subfig4}
 }
\quad
\subfigure[$~$Schrodinger Cat state. The difference between ${\cal B}$ and ${\cal N}$ becomes constant at the large $\alpha$.]{
   \includegraphics[scale =0.42] {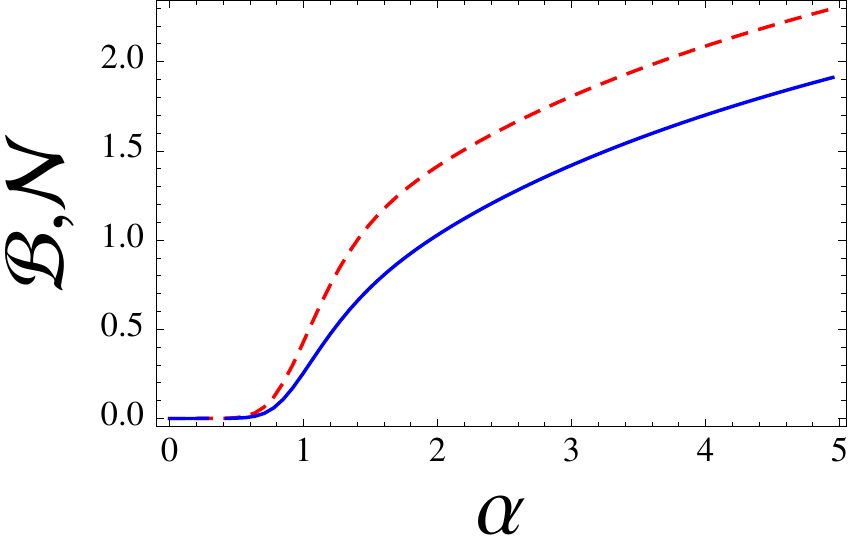}
   \label{fig:subfig3}
 }
\end{center}
\label{myfigure}
\caption{The change of uncertainty are plotted as the parameters of the states are increased. The uncertainty bound ${\cal B}$ (red dashed) are always larger than quantum non-gaussianity ${\cal N}$ (blue line) although their behavior at the asymptotic limit are differed by the state profiles.}
\end{figure}
Formulating entropic UR in Eq.~\eqref{eq:Neg_bound} raises an interesting question: what is the maximum value of ${\cal N}$ achievable for a given value of the uncertainty product $\sigma_x\sigma_p$? In what follows, we shall present some typical examples of non-Gaussian states of a single-mode CV system, and study their neg-entropy ${\cal N}$ against the upper bound $${\cal B}\equiv\ln(2\sigma_x\sigma_p)$$ which is provided in Eq.(\ref{eq:Neg_bound}). The choice of pure states is dictated by the fact that the entropies $H(X)$ and $H(P)$ are strictly concave functions of the state. Hence, the entropic UR as well as its reformulation \eqref{eq:Neg_bound} can only be saturated by pure gaussian states. In several of the examples of non-gaussian (pure and mixed) states to follow, we make use of the annihilation operator defined as
$\hat a=(\hat x+i\hat p)/{\sqrt2}$,
with respect to a vacuum state $\ket0$ such that $\hat a\ket0=0$.

\subsection{Number states} 
As a solution of one dimensional Harmonic oscillator, the number state is defined as $\ket n=(n!)^{-1/2}(\hat a^\dagger)^n\ket0$, and has a wavefunction in configuration space
\begin{equation}
	\psi_n(x)=\sprod{x}{n}=\frac{1}{\sqrt{\sqrt{\pi}2^nn!}}e^{-\frac{x^2}{2}}H_n(x),\label{fockpsi}
\end{equation}
where $H_n(x)$ is the $n$-th Hermite polynomial. The standard deviations are easily found as
\begin{align}
	\sigma_x=\sigma_p=\sqrt{n+1/2}.
\end{align}
On the other hand, the entropies $H(X), H(P)$ for number state involve complicated integrals in general that we do not attemp to compute analytically here. Instead, the numerical evalutation of the functions is demonstrated in Fig.\ref{fig:subfig1}. In that case, one can find that the gaussian distance ${\cal N}$ and its upper bound ${\cal B}$ are both monotonically increasing while the gap between ${\cal N}$ and ${\cal B}$ also increasing. It means that, for the case of number state, the modified uncertainty is not fully saturated by the quantified gaussian distance although the bound has been improved with non-Gaussianity measure.

\subsection{A state with Poisson (or Laplace) distribution}
As a second example of a non-gaussian state, we consider a state whose wavefunction has Poisson (or Laplace) distribution. The functional form of the distribution is
\begin{equation}
	\psi(x)=\sqrt{\frac{2}{\lambda}}e^{-\lambda|x|/2}
\label{eq:Poiss}
\end{equation}
where we set the shape of distribution symmetrically with respect to $x$-axis in order to make its mean value to be trivial. A typical example of the state with Poissonian distribution is coherent state whose photon number distribution is  given as $p(n)=\bar{n}^n e^{-\bar{n}}/n!$. However, we do not concentrate on a particular quantum state but any non-gaussian state with the distribution provided in Eq.(\ref{eq:Poiss}). After fourier transformation of $\psi(x)$, one can also obtain the Cauchy-Lorentz distribution in the conjugate space  as $\varphi(p)=2 \lambda^{3/2}/\sqrt{\pi}(4p^2+\lambda^2)$. With all the distributions, the functions yield
\begin{eqnarray}
&\sigma_x=\sqrt2/\lambda,
&\sigma_p=\lambda/2,\nonumber\\
&H(X)=1-\ln(\lambda/2),
&H(P)=\ln(4\pi\lambda)-2,\\
&{\cal B}=\ln2/2\simeq0.346,
&{\cal N}=1+\ln(e/4\sqrt2)\simeq0.267\nonumber
\end{eqnarray}
where it can be noticed that both the total neg-entropy and its bound are independent of $\lambda$. Therefore, the state with the wave function of Poisson distribution has uniform distribution over the quantum uncertainty relation as well as the non-Gaussianity (gaussian distance).

\subsection{States with photon addition to coherent and squeezed vacuum inputs}
As another example of pure non-gaussian state, we consider states which have been known for single photon added state to coherent and squeezed states. Those states are of high interest because it is an experimentally realizable non-Gaussian states whose operations are fundamental ingredients for continuous variable quantum technologies \cite{Parigi07}. In fact, besides enhancing some existing protocols, it is known that they are able to perform important communication tasks (like entanglement distillation), which are impossible with gaussian states and gaussian operations only. The definition of photon added coheret state is $\ket\psi\propto\hat a^\dagger \hat{D}(\alpha)|0\rangle$ with the displacement operator $\hat{D}(\alpha)=\exp\left(\alpha\hat a^{\dagger}-\alpha^*\hat a\right)$. Without loss of generality, $\alpha$ can be set as a real number within our discussion and the normalized wavefunction reads
\begin{equation}
	\psi_{\alpha}(x)=\frac{e^{-\frac{1}{2} (x-{\bar x})^2} (2 x-\bar{x})}{\sqrt[4]{\pi } \sqrt{\bar{x}^2+2}},
\end{equation}
when $\bar x\equiv\sqrt2\alpha$. Means and variances are easy to be evaluded and they become 
\begin{equation}
\sigma_x=\sqrt{\frac{\bar x^4+12}{2(\bar x^2+2)^2}}~~\mbox{and}~~
\sigma_p=\sqrt{\frac{\bar x^2+6}{2\bar x^2+4}}.
\end{equation}
In the mean time, $H(X)$ and $H(P)$ cannot be integrated analytically in general so that it is numerically evaluated. Fig.\ref{fig:subfig4} shows the asymtoptic behavior of ${\cal N}$ and ${\cal B}$ as the value $\alpha$ is increased. When $\alpha=0$, the state becomes Fock state with single excitation and, in the limiting case $\alpha\rightarrow\infty$, the state approaches to the coherent state $|\alpha\rangle$. Consistantly, the non-gaussianity ${\cal N}$ is changed from the value for a single photon Fock state to a Gaussian state, that is trivial. As it can be found in the figure, the non-gaussianity and the bound  are monotonically decreasing with respect to $\alpha$ while ${\cal N}<{\cal B}$ for any $\alpha$.

For the case of single photon added squeezed vacuum state, the state can be obatined as $\ket\psi\propto\hat a^\dagger\hat{S}(\xi)|0\rangle$ where $\hat{S}(\xi)=\exp\left(\xi\hat a^{\dagger2}-\xi^*\hat a^2\right)$. Wavefunction of the state in the quadrature space reads:
\begin{align}
	\psi_{sq}(x)=\frac{\sqrt{2} x e^{-\frac{x^2}{2 \xi ^2}}}{\sqrt[4]{\pi } \xi ^{3/2}},
\end{align}
where $\xi>0$ is a real parameter which characterizes the degree of squeezing. As a matter of simplicity, we have considered squeezing along the $x$ axis only. In that case, we have
\begin{eqnarray}
&\sigma_x=\sqrt{\frac32}\xi,
&\sigma_p=\sqrt{\frac32}\frac1\xi,\nonumber\\
&H(X)=\ln\sqrt{4\pi\xi^2/e}+\gamma_E,
&H(P)=\ln\sqrt{4\pi/\xi^2e}+\gamma_E,\\
&{\cal B}=\ln3\simeq1.099,
&{\cal N}=2-2 \gamma_E-2 \coth ^{-1}7\simeq0.56,\nonumber
\end{eqnarray}
where $\gamma_E\simeq0.577$ is the Euler-Mascheroni constant. The state has non-trivial constant values ${\cal N}$ and ${\cal B}$ irrespective of squeezing parameter $\xi$ and the values coincide to the one for the single photon Fock state. It instructively indicates that the squeezing operation does not influance the amount of non-gaussianity ${\cal N}$ as well as its upper bound ${\cal B}$.

\subsection{Coherent superposition state: Schr\"odinger cat state}
It is well-known that continuous variable coherent superposition state (also known as Schr\"odinger cat state) is a useful resource for the various purposes of quantum information processing, {\it e.g.}\cite{Ralph03}. In this subsection, we investigate the properties of the coherent superposition (CS) state which are related with the uncertainty relation. The state is defined in general as
\begin{equation}
\ket{\psi}=\frac{1}{\sqrt{2(1+e^{-2|\alpha|^2}\cos\theta )}}(\ket\alpha+e^{i\theta}\ket{-\alpha}),	
\end{equation}
where the coherent state $\ket\alpha$ is implicitly defined as $\hat a\ket\alpha=\alpha\ket\alpha$. We shall restrict to $\alpha\in\mathbb R$ for simplicity, such that the configuration space wavefunction reads
\begin{equation}
	\psi(x)=\frac{\pi^{-1/4}}{\sqrt{2(1+e^{-2|\alpha|^2}\cos\phi )}}[e^{-\frac12(x-\bar x)^2}+e^{i\theta}e^{-\frac12(x+\bar x)^2}]
\end{equation}
where $\bar x=\sqrt2\alpha$. From this, we calculate all quantities of interest (see Fig.\ref{fig:subfig3} for some examples). In the limit of large $|\alpha|$, we can extract some simple analytical expressions. For $\alpha\gg1$ we obtain the asymptotic expressions, which do not depend on $\theta$:
\begin{eqnarray}
&\sigma_x\simeq\sqrt{\frac12+2\alpha^2},
&\sigma_p\simeq\frac{1}{\sqrt{2}},\nonumber\\
&H(X)\simeq\ln\sqrt{4\pi e},
&H(P)\simeq\ln\sqrt{\pi e},\\
&{\cal B}\simeq\ln\sqrt{1+4\alpha^2},
&{\cal N}\simeq\ln\sqrt{\frac14+\alpha^2}.\nonumber
\end{eqnarray}
We can see that ${\cal B}-{\cal N}$ tends to a constant as $\alpha\to\infty$:
\begin{equation}
	{\cal B}-{\cal N}=\ln\sqrt{\frac{1+4\alpha^2}{1/4+\alpha^2}}\simeq\ln2~(\mbox{at}~\alpha\to\infty).
\end{equation}
This is indeed expected from the structure of CS states for large real $\alpha$: the $\hat x$-probability distribution tends to a bimodal one, while the $\hat p$ distribution to a gaussian with variance $1/2$. Hence, as $\alpha$ is increased the sum $H(X)+H(P)$ eventually saturates, while the standard deviation $\sigma_x$ will increase with $\alpha$ linearly (approximately). 

It is instructive to know that ${\cal B}-{\cal N}$ has asymtotic constant value $\ln 2$. It is compatible to the case of coherent state $|\alpha\rangle$ whose variances are constant in every quadratures as $\Delta x_{\theta}=1/\sqrt{2}$. Consequence of the finite variance for coherent state is that the state is viewed as a single point in the large photon number limit as like a point of classical Harmonic oscillator in phase space. Similary, the effect of finite ${\cal B}-{\cal N}$ is to be considered as $({\cal B}-{\cal N})/{\cal B}$ becomes arbitrary small in the large ${\cal B}$ limit. It implies that, in the large ${\cal B}$ scale, the CS state can be viewed as a state whose uncertainty ${\cal B}$ is saturated to the value of non-gaussianity (or gaussian distance) ${\cal N}$ as like coherent field in the phase space. In that regards, it is sensible to argue that the CS state (or Schr\"odinger cat state) is asymtotically optimized with respect to our refined uncertainty relation (\ref{eq:Neg_bound}).

\begin{figure}[t]
\begin{center}
\subfigure[$~$Entropic uncertainty and upper bound]{
   \includegraphics[scale =0.37] {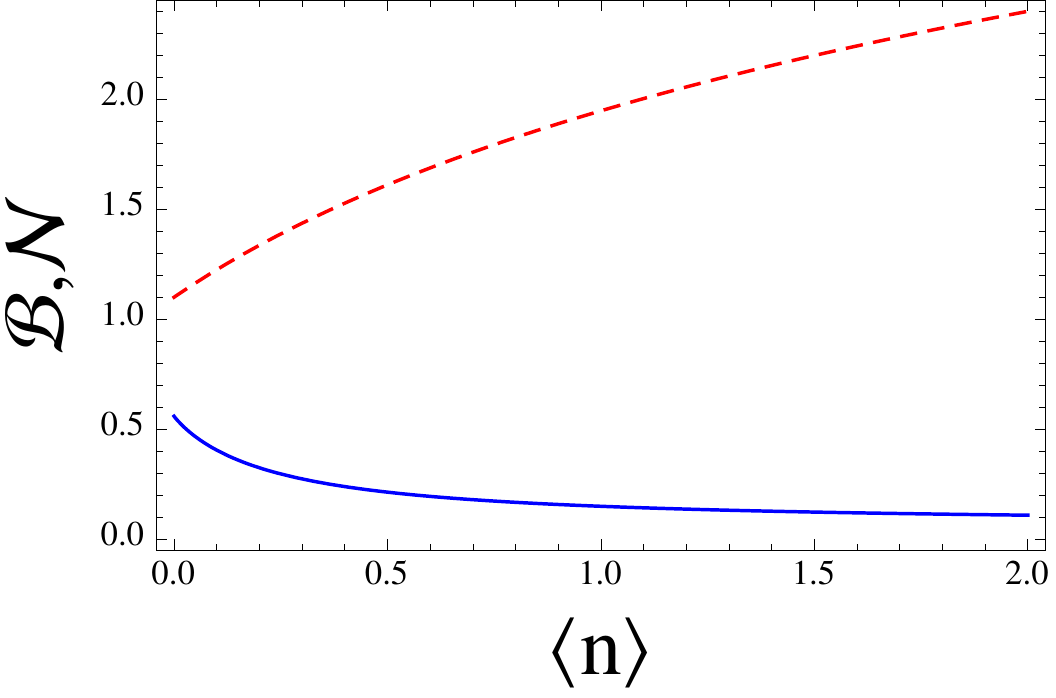}
   \label{fig:thermal_subfig1}
 }
\quad
\subfigure[$~$ The upper bound with purity correction]{
   \includegraphics[scale =0.37] {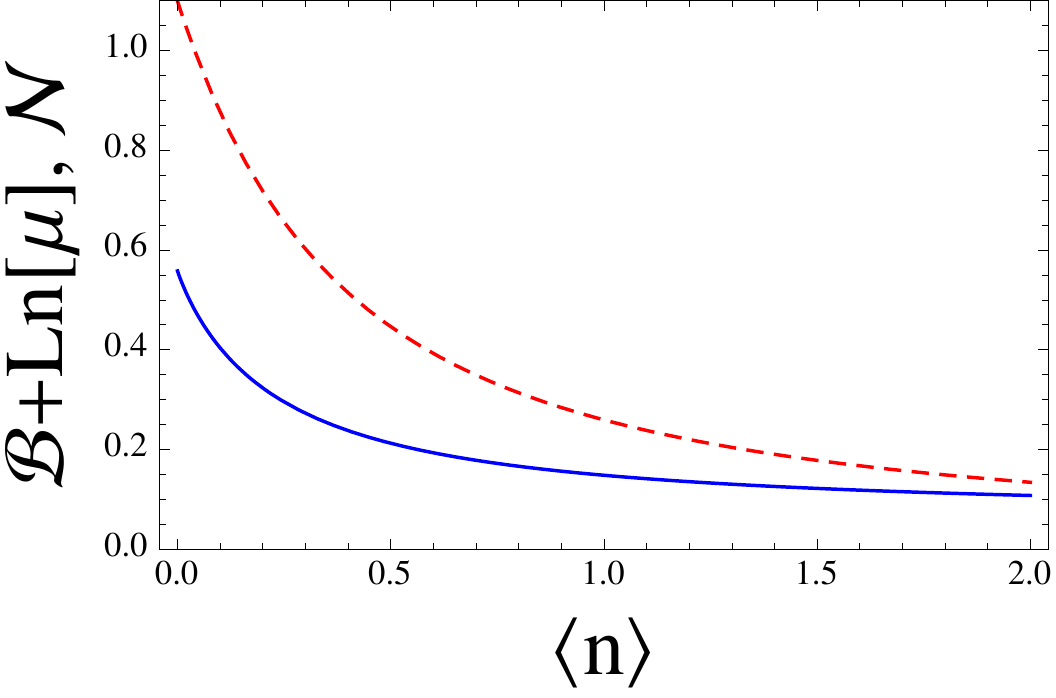}
   \label{fig:thermal_subfig2}
 }
\end{center}
\label{fig:mixed}
\caption{Mixed state example of uncertainty relation. The factors, ${\cal N}$ (Blue line) and ${\cal B}$ (Red dots), in the entropic uncertainty relation are plotted for the case of photon added thermal state. The relation becomes tighter with purity correction.}
\end{figure}

\subsection{Photon added thermal state; Non-gaussian mixed state}
Upto this point, we had investigated the properties of entropic uncertainty relation for the case of  non-gaussian pure states. In a generalized case, it is still intriguing whether a non-gaussian {\it mixed state} follows the analytic behavior of the modified uncertainty relation. Inspired by the recent results \cite{Dodonov02,Mandilara12}, it can be conjectured that the lower bound of general entropic UR contains additional term with mixedness and it reads 
\begin{equation}
\label{eq:Neg_bound_mixed}
{\cal J}(X)+{\cal J}(P) \leq \ln(2\sigma_x\sigma_p)+\ln(\mu)
\end{equation}
with purity of the state defined as $\mu=\mbox{Tr}[\rho^2]$. It is worth to mention that modification of CV entropic UR with purity is quantativiely different from the case of finite dimensional entropic UR. The difference can be identified from the fact that the additional term by mixedness in the finite dimensional case is given by the von Neumann entropy $S(\rho)$ \cite{Korzekwa14}. For the case of CV state, the inequality relation Eq.(\ref{eq:Neg_bound_mixed}) becomes equality  in the gaussian limit which is derived simply from the modified minimum Heisenberg uncertainty relation \cite{Dodonov02}. However, the inequality is nontherless trivial when the given state has general non-gaussian profile and it should be stressed that validity of the relation with purity $\mu$ for a general state has not been genuinely confirmed before. Here we provide heuristic aruguement of the modified inequality (\ref{eq:Neg_bound_mixed}) through a specific example of non-gaussian mixed state.

In order to analyzing the uncertainty relation (\ref{eq:Neg_bound_mixed}), we consider a photon added thermal state 
\begin{equation}
\rho_{add}=\hat{a}^{\dagger}\rho_{th}\hat{a}~~~\mbox{where}~~~\rho_{th}=\frac{1}{\bar{n}+1}\sum \left(\frac{\bar{n}}{\bar{n}+1}\right)^n|n\rangle\langle n|
\end{equation}
and $\bar{n}$ is mean photon number of the state. It can be taken as a simple example of non-gaussina mixed state. Probability distribution of the state in the quadrature space $p_{\bar{n}}(x)=\langle x|\rho_{add}|x\rangle$ is given as
\begin{equation}
\label{eq:thermal_added}
p_{\bar{n}}(x)=|\psi(x)|^2=\frac{\bar{n}(1+2\bar{n})+2(1+\bar{n})x^2}{\sqrt{\pi}(1+2\bar{n})^{5/2}} e^{-\frac{x^2}{1+2\bar{n}}}
\end{equation}
and it can be used to evaluate the relavent quantity for URs. Figures in (\ref{fig:thermal_subfig1}) and (\ref{fig:thermal_subfig2}) are plot of neg-entropies ${\cal N}$ for the photon added thermal state and their upper bounds ${\cal B}$, with and without purity term $\ln(\mu)$. The validation of result can be made at the extreme limit such that $\bar n=0$ makes the state pure where the purity term disappears and $\bar n\rightarrow\infty$ leads the state approached to a completely mixed gaussian state whose bound coincides to the value of non-Gaussianity. It means that entropic UR becomes tigher with log scaled purity term simliar to Heisenberg UR.

\begin{figure}[t]
	\begin{center}
		\includegraphics[width=.7\linewidth]{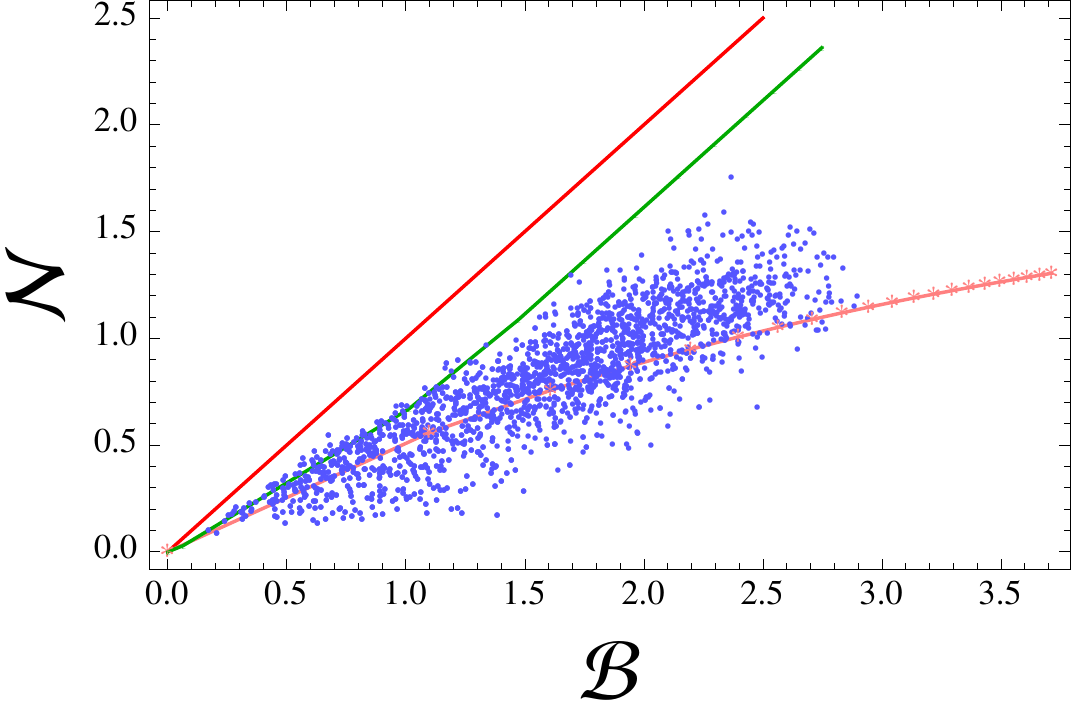}
	\end{center}
\caption{Comparison of the total neg-entropy ${\cal N}$ versus the state-dependent bound ${\cal B}$ for different examples of quantum states together with the randomly generated pure state. It shows that the coherent superposition state behaves closely to the lower bounds while there exists more optimal state.}
\label{random}
\end{figure}

\section{Quantative analysis for optimal uncertainty with pseudo random states }
Here, we ask a question whether there exists a class of non-gaussian state which saturates entropic UR in Eq. (\ref{eq:Neg_bound}) completely. Such a non-gaussian state will attain the minimum entropic bound and show quantatively different behavior compared to the gaussian minimum uncertainty states ({\it e.g.} coherent state). It means that the minimal entropic UR state can contain maximum information from the information theoretical point of view. So far, it turns out that none of the states in examples above do not satisfy the minimum entropic uncertainty principles while they still do not exclude potential existance of a non-gaussian minimum uncertainty state.

In order to investigate the existance of the state with optimal uncertainty, we try to locate the state using random state generation method. Within an appropriate numerical complexity, we generate 11 pseudo-random complex numbers $c_0,...,c_{10}$, and construct a pure quantum state. Then, a set of states are obtained as
\begin{equation}
	\ket\psi=\frac1{\sqrt{\sum_n|c_n|^2}}\sum_{n=0}^{10}c_n\ket{n}.
\label{eq:random}
\end{equation}
The wavefunction in configuration space $\psi(x)=\langle x|\psi\rangle$ can be found easily by exploiting Eq.~\eqref{fockpsi}, and all relevant statistical properties can be calculated numerically from $p(x)=|\psi(x)|^2$. The state is constructed to test the existance of the optimal entropic UR state in comparison to the non-gaussian states that we had investigated in the privious section. 

The figure (\ref{random}) shows the comparison between the total neg-entropy ${\cal N}$ versus the state-dependent bound ${\cal B}$ for different examples of quantum states together with the randomly generated pure state in Eq. (\ref{eq:random}). In the figure, red line represents the minimal bound of entropic UR when a set of non-gaussian states satisfy ${\cal N}={\cal B}$. It means that any physical state cannot go beyond the region above the red line. It also shows that the coherent superposition state (green line) behaves closely to the entropic lower bounds while the number state (pink dots) is deviated from minimal uncertainty line at a larger non-gaussian region ${\cal B}\gg1$. 

Blue dots represent ${\cal N}/{\cal B}$ ratio of the psudo-random state and they are scattered around region below the minimal uncertainty bound. All the states exist in the physical region as it is found. Additionally, it can be found that there are states that go beyond the green line for coherent superposition state although their deviation from the state are not very huge. It means that there are states that is more optimal than coherent superposition state in terms of entropic UR. In this investigation, we can predict the existance of optimal entropic UR state although it is not definitive whether there can be any analytic expression of a state which satisfy ${\cal N}={\cal B}$. Once more, together with all the results that we have so far, we can conclude that there can be a class of state which satisfies ${\cal N}={\cal B}$ and the state is information theoretically optimized together with our new characterization of non-gaussianity of a quantum state.

\section{Remarks}
In this work, we have shown that the maximum entropy principle by Shannon allows one to rewrite the entropic UR in a refined way as it coincides with modified Heisenberg's uncertainty relation. From the derivation, a new characterization of a quantum non-Gaussianity has been followed from the fact that the principle explicitly features deviation of the position and the momentum probability densities from appropriate reference Gaussians simutaneously. The non-Gaussianities and its upper bounds of various non-gaussian CV states have been evaluated and it turns out that coherent superposition state asymptotically optimizes the refined entropic UR. Together with non-Gaussianity, when the state is in a statistical mixture, it is also found that lower bound of entropic uncertainty relation is enhanced by purity. A potiential existance of completely optimal entropic UR state has been investigated through the random state generation method. As its results, a state closer to the mininal uncertainty bound than CS state is identified in the small uncertainty, ${\cal B}$, region. An operational meaning of the new class of information theoretically optimized state is left as its future studies.

\section{Acknowledgments}
The author acknowledges Dr. Tommaso Tufarelli and Prof. M.S. Kim for their extensive discussion on the topic. I also would like to thank people who help me during my stay in Oxford. This work was done with support of Oxford Martin School, ICT R\&D program of MSIP/IITP (No.2014-044-014- 002) and National Research Foundation (NRF) grant (No.NRF-2013R1A1A2010537) funded by Korean Government.

\end{document}